\documentclass[prl,twocolumn,superscriptaddress]{revtex4-1}
\usepackage{bm}
\usepackage{bbm}
\usepackage{multirow}
\usepackage{float}
\usepackage{natbib}
\usepackage{CJK}
\usepackage{color}
\usepackage[usenames,dvipsnames]{xcolor}
\usepackage[version=3]{mhchem} 
\usepackage[T1]{fontenc}       
\usepackage[export]{adjustbox} 

\usepackage[a4paper,colorlinks=true,
linkcolor=blue,citecolor=blue,
pdfauthor={ },
pdftitle={ },
pdfsubject={ },
pdfkeywords={ }]{hyperref}

\begin{document}

\title{Single DNA Electron Spin Resonance Spectroscopy in Aqueous Solutions}


\author{Fazhan Shi}
\altaffiliation{These authors contributed equally to this work.}
\affiliation{CAS Key Laboratory of Microscale Magnetic Resonance and Department of Modern Physics, University of Science and Technology of China (USTC), Hefei 230026, China}
\affiliation{Hefei National Laboratory for Physical Sciences at the Microscale, USTC}
\affiliation{Synergetic Innovation Center of Quantum Information and Quantum Physics, USTC}

\author{Fei Kong}
\altaffiliation{These authors contributed equally to this work.}
\affiliation{CAS Key Laboratory of Microscale Magnetic Resonance and Department of Modern Physics, University of Science and Technology of China (USTC), Hefei 230026, China}
\affiliation{Hefei National Laboratory for Physical Sciences at the Microscale, USTC}
\affiliation{Synergetic Innovation Center of Quantum Information and Quantum Physics, USTC}

\author{Pengju Zhao}
\altaffiliation{These authors contributed equally to this work.}
\affiliation{CAS Key Laboratory of Microscale Magnetic Resonance and Department of Modern Physics, University of Science and Technology of China (USTC), Hefei 230026, China}
\affiliation{Synergetic Innovation Center of Quantum Information and Quantum Physics, USTC}

\author{Xiaojun Zhang}
\affiliation{Department of Chemistry, University of Southern California, Los Angeles, CA 90089, USA}

\author{Ming Chen}
\affiliation{CAS Key Laboratory of Microscale Magnetic Resonance and Department of Modern Physics, University of Science and Technology of China (USTC), Hefei 230026, China}
\affiliation{Synergetic Innovation Center of Quantum Information and Quantum Physics, USTC}

\author{Sanyou Chen}
\affiliation{CAS Key Laboratory of Microscale Magnetic Resonance and Department of Modern Physics, University of Science and Technology of China (USTC), Hefei 230026, China}
\affiliation{Synergetic Innovation Center of Quantum Information and Quantum Physics, USTC}

\author{Qi Zhang}
\affiliation{CAS Key Laboratory of Microscale Magnetic Resonance and Department of Modern Physics, University of Science and Technology of China (USTC), Hefei 230026, China}
\affiliation{Synergetic Innovation Center of Quantum Information and Quantum Physics, USTC}

\author{Mengqi Wang}
\affiliation{CAS Key Laboratory of Microscale Magnetic Resonance and Department of Modern Physics, University of Science and Technology of China (USTC), Hefei 230026, China}
\affiliation{Synergetic Innovation Center of Quantum Information and Quantum Physics, USTC}

\author{Xiangyu Ye}
\affiliation{CAS Key Laboratory of Microscale Magnetic Resonance and Department of Modern Physics, University of Science and Technology of China (USTC), Hefei 230026, China}
\affiliation{Hefei National Laboratory for Physical Sciences at the Microscale, USTC}
\affiliation{Synergetic Innovation Center of Quantum Information and Quantum Physics, USTC}

\author{Zhecheng Wang}
\affiliation{CAS Key Laboratory of Microscale Magnetic Resonance and Department of Modern Physics, University of Science and Technology of China (USTC), Hefei 230026, China}
\affiliation{Synergetic Innovation Center of Quantum Information and Quantum Physics, USTC}

\author{Zhuoyang Qin}
\affiliation{CAS Key Laboratory of Microscale Magnetic Resonance and Department of Modern Physics, University of Science and Technology of China (USTC), Hefei 230026, China}
\affiliation{Synergetic Innovation Center of Quantum Information and Quantum Physics, USTC}

\author{Xing Rong}
\affiliation{CAS Key Laboratory of Microscale Magnetic Resonance and Department of Modern Physics, University of Science and Technology of China (USTC), Hefei 230026, China}
\affiliation{Hefei National Laboratory for Physical Sciences at the Microscale, USTC}
\affiliation{Synergetic Innovation Center of Quantum Information and Quantum Physics, USTC}

\author{Jihu Su}
\affiliation{CAS Key Laboratory of Microscale Magnetic Resonance and Department of Modern Physics, University of Science and Technology of China (USTC), Hefei 230026, China}
\affiliation{Hefei National Laboratory for Physical Sciences at the Microscale, USTC}
\affiliation{Synergetic Innovation Center of Quantum Information and Quantum Physics, USTC}

\author{Pengfei Wang}
\affiliation{CAS Key Laboratory of Microscale Magnetic Resonance and Department of Modern Physics, University of Science and Technology of China (USTC), Hefei 230026, China}
\affiliation{Hefei National Laboratory for Physical Sciences at the Microscale, USTC}
\affiliation{Synergetic Innovation Center of Quantum Information and Quantum Physics, USTC}

\author{Peter Z. Qin}
\email{pzq@usc.edu}
\affiliation{Department of Chemistry, University of Southern California, Los Angeles, CA 90089, USA}

\author{Jiangfeng Du}
\email{djf@ustc.edu.cn}
\affiliation{CAS Key Laboratory of Microscale Magnetic Resonance and Department of Modern Physics, University of Science and Technology of China (USTC), Hefei 230026, China}
\affiliation{Hefei National Laboratory for Physical Sciences at the Microscale, USTC}
\affiliation{Synergetic Innovation Center of Quantum Information and Quantum Physics, USTC}

\maketitle

{\textbf{
Magnetic resonance spectroscopy of single biomolecules under near-physiological conditions may substantially advance understanding of biological function, yet remains very challenging. Here we use nitrogen-vacancy centers in diamonds to detect electron spin resonance spectra of individual, tethered DNA duplexes labeled with a nitroxide spin label in aqueous buffer solutions at ambient temperatures. This paves the way for magnetic resonance studies on single biomolecules and their inter-molecular interactions in a native-like environment.}}

\begin{figure*}[ht]
\includegraphics[width=2\columnwidth, left]{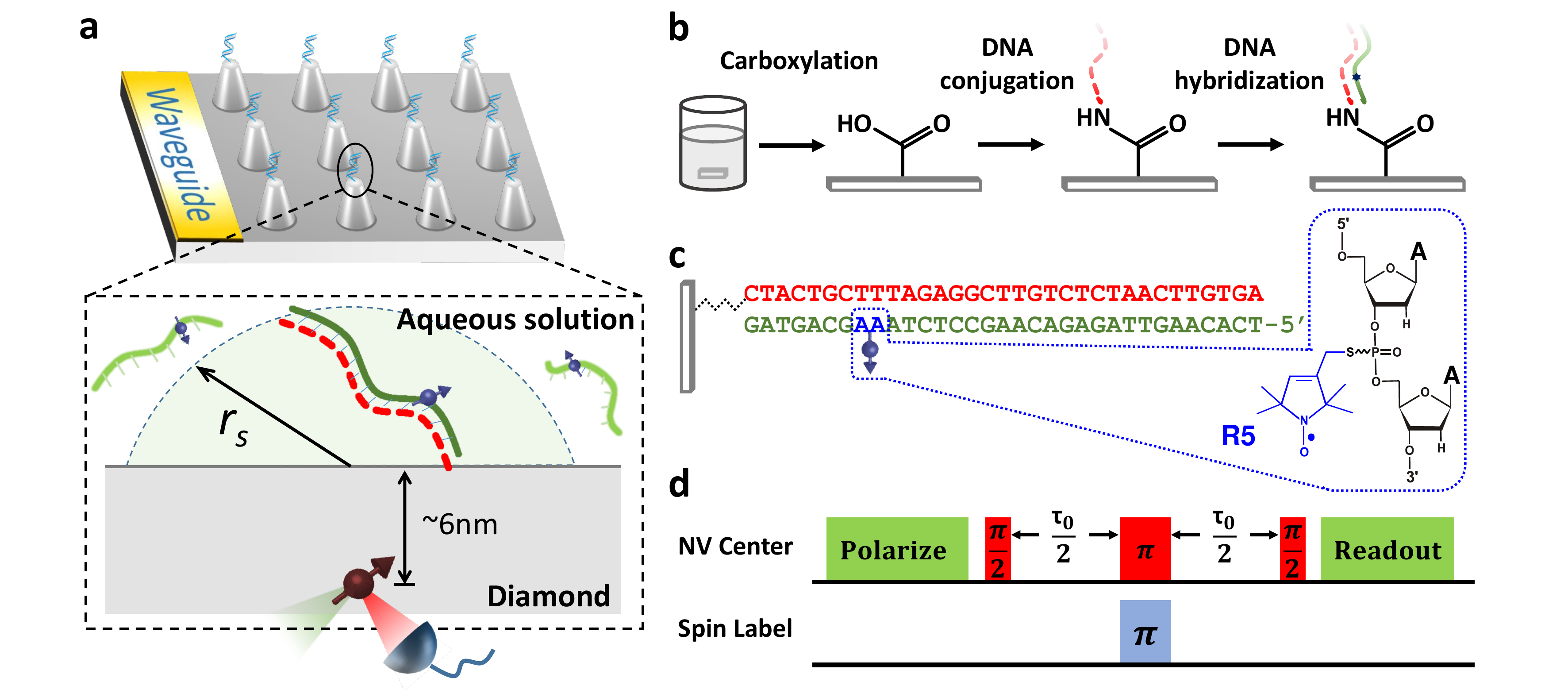}
    \caption{\textbf{Experimental setup for NV-detection of single spin-labeled DNAs under physiological conditions. }
   (\textbf{a})
   Schematic of the diamond pillars with surface-tethered DNA. The inset illustrates that hybridizing a DNA strand (green solid line) with a spin-label (blue arrow) to an unlabeled tethered strand (red dashed line) localizes a spin-labeled duplex within the detection volume of an NV sensor (red arrow).
   (\textbf{b}) The chemical tethering scheme. A 5'-NH$_2$ modified DNA strand (red line, top) was reacted to the carboxylate groups at the diamond surface. The complementary strand (green line, bottom) was then captured by hybridization to the tethered DNA.
   (\textbf{c}) The tethered DNA duplex with a R5 label (blue arrow) 8-base-pair away from the linker. The right inset shows the R5 structure with either $^{14}$N or $^{15}$N at the pyrroline ring.
   (\textbf{d}) Double electron-electron resonance pulse sequence for target spin detection.
 }
    \label{Fig:setup}
\end{figure*}

Given the enormous power of magnetic resonance spectroscopy, including Nuclear Magnetic Resonance (NMR) and Electron Spin Resonance (ESR),  in interrogating molecular structure and dynamics, intensive efforts are being devoted to developments of single spin magnetic resonance spectroscopy, and rapid progresses are being made \cite{Wrachtrup2016JMR}.  The use of nitrogen-vacancy (NV) centers in diamonds is one of the most promising venues for single spin detection \cite{Wrachtrup2008Nature, Lukin2008Nature, Degen2017RMP},  for example, we have previously reported single-molecule ESR spectroscopy of proteins labeled with a nitroxide spin label and embedded in a poly-lysine layer at diamond surfaces \cite{Du2015Science}. However, in all prior NV work, the target is either embedded within the diamond lattice \cite{Shi2013PRB,Yacoby2013NatPhys, Yacoby2014nnano} or fixed at the diamond surface \cite{Wrachtrup2011NJP,Lukin2014NanoLett, Du2015Science}.
A majority of biomolecules function in aqueous solutions under ambient temperatures (i.e., physiological conditions), at which the molecules undergo a high degree of motions. NV-detection of single molecules at physiological conditions posts significant additional challenges as compared to studies at a stationary solid phase, and has not yet been reported.

We reported here two technical advancements to enable ESR measurements of single spin-labeled DNAs at physiological conditions. We implemented a diamond pillar array design (\textbf{Fig.~\ref{Fig:setup}a}, \textbf{Supplementary Note 1} and Online Methods), which reduces the detection time by nearly one order.
This enabled one to record multiple ESR spectra before the labels are quenched due to laser irradiation \cite{Wrachtrup2017SciAdv}.
 Furthermore, to confine a spin-labeled DNA duplex within the $\sim$10 nm detection range from a shallowly embedded NV, a chemical tethering scheme was devised \cite{Lukin2016Science}, in which a non-labeled DNA was covalently attached to the diamond surface, then hybridized with a complementary strand with a spectroscopic label (\textbf{Fig.~\ref{Fig:setup}b} and Online Methods). Atomic Force Microscopy imaging indicated that the tethered DNA forms an evenly distributed single layer at the diamond surface (\textbf{Supplementary Note 2A}). Characterizations using a Cy3-labeled DNA complementary strand and fluorescence confocal microscopy demonstrated that DNA duplex localization depends on the presence of the covalently attached non-labeled strand. Under experimental conditions used in this work, spacing between the DNA duplexes was estimated to be 21 nm (\textbf{Supplementary Note 2B}). This resulted in a 14\% probability of a single DNA duplex located within the detection range of an NV sensor, while the probability of detecting two or more DNA duplexes was 1\% (\textbf{Supplementary Note 3B}). As such, the signal detected from a single NV center was predominately from a single DNA duplex.

\begin{figure}
\includegraphics[width=1.0\columnwidth, left]{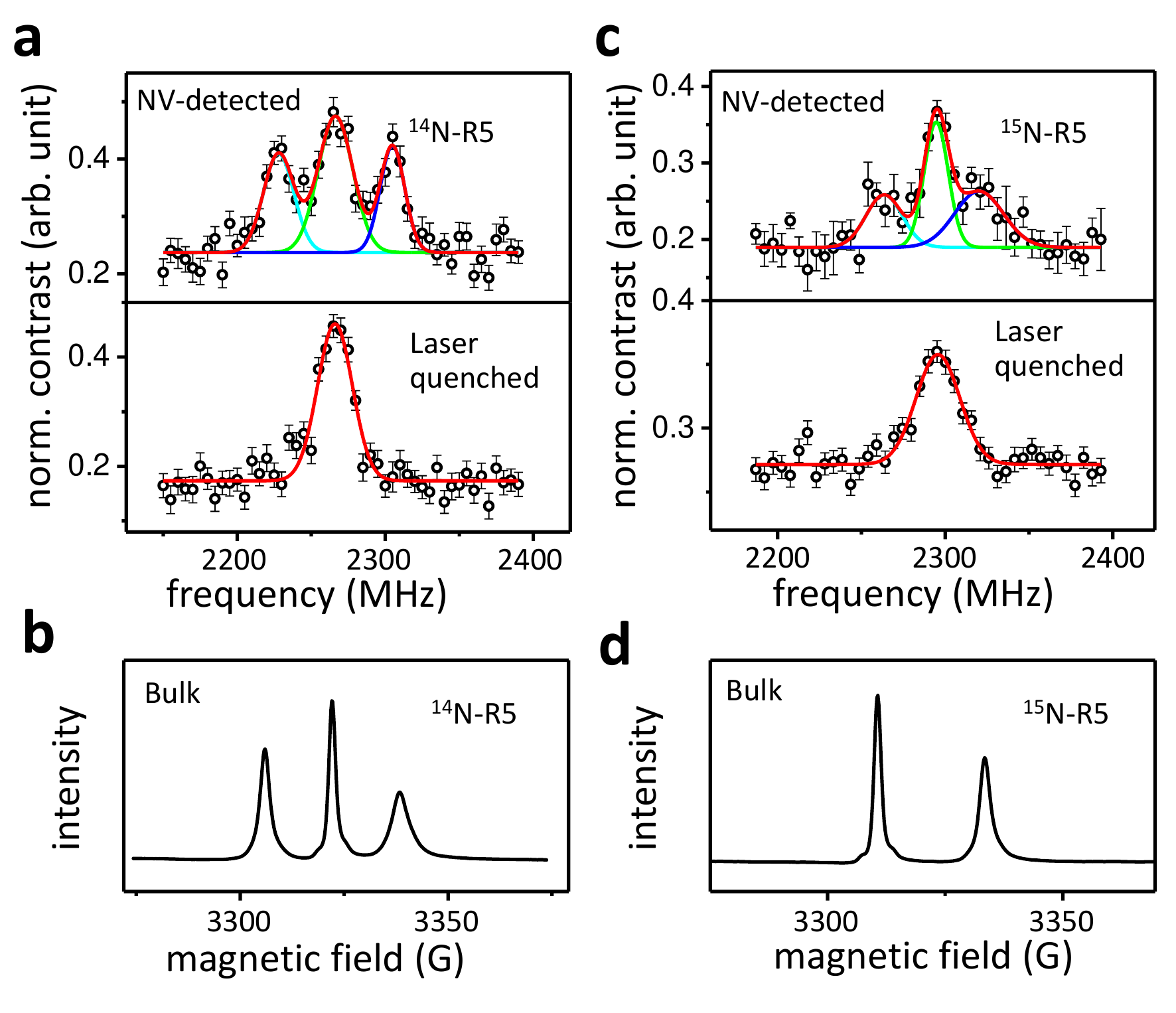}
    \caption{\textbf{ESR spectra detected at single-molecule and bulk level.}
   (\textbf{a}) NV-detected spectrum of a single $^{14}$N-R5 labeled DNA (circles: experiment; lines: Gaussian peak fit). Top panel: an earlier time-point spectrum with three peaks.
   The two side-peaks ($2228.0\pm1.3$ and $2304.8\pm1.0$ MHz) give a measured $2A_{\text{iso}}=76.8\pm2.3$ MHz. The center peak ($2266.4\pm1.0$ MHz) includes signals from the $^{14}$N-R5 and paramagnetic surface defects. Bottom panel: a spectrum observed after hours of laser irradiations that quench the nitroxide. Only one peak ($2266.4\pm1.0$ MHz) corresponding to the defects is present. Error bars indicate $\pm$1 standard error of the mean (s.e.m) and the data shown were obtained from measurements repeated 0.8 and 1.2 million times, respectively, for the top and bottom panel.
   (\textbf{b}) Ensemble X-band spectrum of the corresponding $^{14}$N-R5 labeled duplex, which shows the three-line pattern expected from hyperfine interaction between the electron spin ($S=1/2$) and the $^{14}$N nucleus ($I=1$) with $2A_{\text{iso}}=90.6$ MHz. 
   (\textbf{c}) NV-detected ESR spectrum of single DNA duplexes labeled with a $^{15}$N-R5. The two side peaks, which were observed initially (top panel) but disappeared within one hour (bottom panel), centered respectively at $2263.5\pm2.5$ and $2320.1\pm5.8$ MHz, yielding an $A_{\text{iso}}=56.6\pm8.3$ MHz. The center peak ($2294.6\pm1.2$ MHz) at both panels represents the surface defects. 
   Error bars indicate $\pm$1 s.e.m, and the data shown were obtained from measurements repeated 4 and 20 million times, respectively, for the top and bottom panel.
   (\textbf{d}) Ensemble X-band spectrum of the corresponding $^{15}$N-R5 labeled duplex, which shows the two-line pattern expected from hyperfine interaction between the electron spin ($S=1/2$) and the $^{15}$N nucleus ($I=1/2$) with $A_{\text{iso}}=64.0$ MHz.
   ~See \textbf{Supplementary Notes 3 $\&$ 4} for additional data and discussions.
 }
    \label{Fig:exp_result}
\end{figure}

Using the tethering scheme, DNA duplexes with a covalently attached nitroxide spin label (designed as R5 \cite{Tangprasertchai2015}, see \textbf{Fig. \ref{Fig:setup}c}) were localized at the diamond pillar surface (\textbf{Fig.~\ref{Fig:setup}b,c}), and  ESR spectrum was detected by a double electron-electron resonance pulse sequence \cite{Du2015Science} (\textbf{Fig.~\ref{Fig:setup}d, Supplementary Note 3A}).
\textbf{Fig.~\ref{Fig:exp_result}a} shows an example of an NV-detected spectrum of a $^{14}$N R5-labeled DNA obtained with an external magnetic field $B_0 = 809$ G, which shows three peaks. The isotropic hyperfine coupling, measured between the two side peaks, was $A_{\text{iso}}^{^{14}\text{N}} = 38.4\pm1.2$ MHz, similar to that obtained from an ensemble measurement (\textbf{Fig.~\ref{Fig:exp_result}b} and \textbf{Supplementary Note 4}) (see further discussion of $A_{\text{iso}}$ below).
Note that while both the NV-detected spectrum (\textbf{Fig.~\ref{Fig:exp_result}a} top) and the ensemble spectrum (\textbf{Fig.~\ref{Fig:exp_result}b}) show a three-line pattern expected due to hyperfine interactions between the electron spin ($S=1/2$) and the $^{14}$N nucleic spin ($I=1$), the center peak of the NV-detected spectrum (gyromagnetic ratio of $2.801\pm0.002$ MHz/G or g-factors = $2.001\pm0.002$) is likely composed of signals from both the $^{14}$N nitroxide and paramagnetic diamond surface defects \cite{Yacoby2014nnano},
 which have similar g-factors and therefore are not separable in the measurements (see also \textbf{Supplementary Note 3C-3E}).
Consistent with this interpretation, the two side-peaks disappeared upon prolong laser irradiation that has been reported to quench the nitroxide (\textbf{Fig.~\ref{Fig:exp_result}a}, bottom)  \cite{Wrachtrup2017SciAdv}.
Furthermore, for diamond surfaces not exposed to spin-labeled DNAs, the NV-detected spectra showed only one peak without splitting (\textbf{Supplementary Note 3D}). 
Together, observations of the side-peaks in the NV-detected spectra indicated that for the first time, single spin-labeled DNA molecules were detected by magnetic resonance in solutions at ambient temperatures.

To further verify detection of nitroxide-labeled DNAs, the $^{14}$N R5 label was substituted to a $^{15}$N R5, which is expected to change the hyperfine coupling (\textbf{Supplementary Note 4}). Indeed, prior to nitroxide decomposition, the NV-detected spectrum showed two side peaks (\textbf{Fig.~\ref{Fig:exp_result}c}, top) that gave an $A_{\text{iso}}^{^{15}\text{N}}=56.6$ MHz.
This ratio of $A_{\text{iso}}^{^{15}\text{N}}/A_{\text{iso}}^{^{14}\text{N}}=1.47\pm0.26$ is consistent with that of the gyromagnetic ratio $|\gamma_{^{15}\text{N}}/\gamma_{^{14}\text{N}}| = 1.40$, thus confirming NV detection of spin-labeled DNA. Note that while the $^{15}$N R5 labeled DNA showed the expected two-line pattern in the ensemble measurements (\textbf{Fig.~\ref{Fig:exp_result}d} and \textbf{Supplementary Note 4}), the NV-detected early time point spectrum shows three peaks (\textbf{Fig.~\ref{Fig:exp_result}c}, top). The center peak was assigned to the paramagnetic diamond surface defects, which is persistent upon prolong laser radiation (\textbf{Fig.~\ref{Fig:exp_result}c}, bottom). Overall, the $^{14}$N and $^{15}$N data unambiguously demonstrate detection of an external nitroxide labeled DNA.

Analyses of NV-detected spectra led to a number of interesting observations. First, the $A_{\text{iso}}$ values varied by $\sim$12\%, reflecting heterogeneity among the individual molecules in the solution (\textbf{Supplementary Note 3C}). All $A_{\text{iso}}$ values measured from NV-detected spectra were smaller ($6\sim16\%$) than that measured in the bulk solution (\textbf{Supplementary Note 3C}).
It is known that a more hydrophobic environment, such as that at the diamond surface, reduces hyperfine couplings \cite{Kurad2003}. the $A_{\text{iso}}$ variations may reflect heterogeneity of the polarity profile at the individual DNA duplexes, although many other factors (e.g., spin label dynamics, local electrostatics) cannot be completely ruled out (see also \textbf{Supplementary Note 5A}).
A second observation is that the NV-detected DNA spectrum showed sharp side-peaks with very small lineshape variations (\textbf{Fig. \ref{Fig:exp_result}} and \textbf{Supplementary Fig. 6}). This differs from those measured from proteins fixed in the poly-lysine layer \cite{Du2015Science}, and reflects the high degree of motions expected from a DNA in solutions (\textbf{Supplementary Note 5A}). Analyses showed that the side-peaks of the NV-detected spectrum were best matched with simulations in which the label undergoes an isotropic rotation with a rotational correlation time $\tau$ of 1.0 ns (\textbf{Supplementary Note 5B}), which is comparable to that reported in bulk solution measurements of R5-labeled DNAs tethered to nano-diamonds \cite{DNA_nanodiamond}.
Overall, these analyses indicate that analyzing the NV-detected single molecule spectrum provides information on the local environment and motional dynamics of the biomolecule. However, a lot more work, including significant increases in the number of spectrum observed, is required in order to retrieve functional information regarding the target biomolecule.

In conclusion, we developed NV-based methods that enabled ESR spectroscopic analyses of single spin-labeled DNA molecules in aqueous solutions at ambient temperatures.
As demonstrated by developments of single-molecule methodologies in other fields such as fluorescent spectroscopy \cite{zhuang2000single} and force microscopy \cite{AFM}, the ability to conduct single molecule measurements in solutions opens up a large number of possibilities for investigating structure, dynamics, and interaction of biomolecules in their native environment.
The work reported represents a significant step forward towards single-molecule magnetic resonance investigation of biomolecular structure and function.

\section{Methods}
Methods, including statements of data availability and any associated accession codes and references, are available in the online version of the paper.

Note: Any Supplementary Information and Source Data files are available in the online version of the paper.

\vspace{1cm}
\section{Acknowledgements}
This work was supported in part by the National Key R\&D Program of China (Grants No. 2016YFA0502400, 2013CB921800), the National Natural Science Foundation of China (Grants No. 81788104, 11227901, 31470835, 91636217, 11722544; PZQ: 21328101), CAS (Grants No. XDB01030400, QYZDY-SSW-SLH004, YIPA2015370), the CEBioM, the Fundamental Research Funds for the Central Universities (WK2340000064, WK2030040088), and the US National Science Foundation (PZQ: CHE-1213673, MCB-1716744).

\section{Author contributions}
J.D. supervised the entire project. J.D., F.S. and P.Z.Q. designed the experiments. F.K., P.Z., and F.S. performed the experiments. X.Z. and P.Z.Q. prepared the DNA duplex. X.Z. M.C. X.R. J.S. and P.Z.Q. and measured the ensemble data. P.Z., M.C., Z.W., S.C., and P.Z.Q. carried the chemical bonding process. M.W., X.Y., and P.W. fabricated the pillar and measured the imaging by AFM. Q.Z. tested the coherence of NVs. F.K. and Z.Q. performed the calculation. F.S., F.K., P.Z., P.Z.Q., and J.D. wrote the manuscript. All authors discussed the results and commented on the manuscript.

\section{Additional information}
Correspondence and requests for materials should be addressed to J.D. and P.Z.Q.
\\
\\
\section{Competing financial interests}
The authors declare no competing financial interests.

\renewcommand\refname{Reference}

\bibliography{DNASolution}

\bibliographystyle{naturemag}

\clearpage

\section{ONLINE METHODS}

\textbf{Diamond Sensors.}
All diamonds used were obtained commercially, 100-oriented, and electronic-grade. NV centers were created by implantation of 4 keV $^{15}$N$_{2}^{+}$ ions with a dose of $1\times10^{11}$ $\text{cm}^{-2}$. The diamond nanopillars were fabricated by electron beam lithography (EBL) and reactive ion etching (RIE) to enhance the photon collection efficiency \cite{Hausmann2011}. First the negative electron-beam resisting hydrogen silsesquioxane (HSQ) was spun on the diamond to a thickness of 350 nm. The HSQ layer was patterned by electron-beam writing followed by tetra-methyl ammonium hydroxide (TMAH, 4\%) developing. The nanopillars were then formed after reactive ion etching with mixed CHF$_3$ and O$_2$, with the etching depth at $\sim$ 400 nm. Finally hydrofluoric acid was used to remove the HSQ resisting layer.

\textbf{DNA samples.}
All DNAs were produced by solid-phase chemical synthesis and obtained commercially. The DNA strand for covalent attachment to diamond surface (i.e., NH$_2$-DNA) has a sequence of 5' NH$_2$-CTACTGCTTTAGAGGCTTGTCTCTAACTTGTGA-3', with "NH$_2$" representing an Amino Modifier C6 at the 5' terminus. The complementary DNA strand (designated as s1) has a sequence of 5'-TCACAAGTTAGAGACAAGCCTCTAAAGCAGTAG-3'. The 5' Cy3-modified s1 strand (Cy3-s1) was obtained commercially. The R5 spin-label [1-oxyl-2,2,5,5-tetramethyl-pyrroline] (\textbf{Fig.~\ref{Fig:setup}c}) was attached to phosphorothioate-modified s1 strands following previously reported protocols and purified by HPLC \cite{Qin2007, Tangprasertchai2015}. The labeling efficiency of the DNA was estimated to be $> 90\%$ by a spin counting procedure \cite{Zhang2009}.

\textbf{Covalent attachment of DNA at the diamond surface.}
Prior to DNA attachment, the diamond surface was cleaned in four steps: (a) submerging in Piranha solution (2:1 mixture of concentrated H$_{2}$SO$_{4}$ and 30\% hydrogen peroxide) at $150 ^{\circ}$C for at least 4 hours; (b) submerging in concentrated HNO$_{3}$ at $90 ^{\circ}$C for one hour; (c) submerging in 1M NaOH at $90 ^{\circ}$C for one hour; and (d) submerging in 1M HCl at $90 ^{\circ}$C for one hour. Following each step, the diamond was rinsed with deionized water.

Following surface cleaning, a freshly prepared solution containing 10 $\mu$M NH$_2$-DNA, 5 mM EDC [1-ethyl-3-(-3-dimethylaminopropyl) carbodiimide hydrochloride, Sigma-Aldrich (39391-50ML)] in 100 mM MES (pH 5.0) was applied to the diamond surface. Reaction was allowed to proceed at room temperature for 30 minutes, after which the diamond surface was rinsed with deionized water. This procedure was repeated for 3 times to maximize the amount of DNA tethered at the diamond surface.

To hybridize the complementary strand to the tethered DNA strand, a 2 $\mu$M Cy3 or R5-labeled s1 strand solution was prepared in 100 mM NaCl buffer. 2 $\mu$L of this solution was added to the diamond surface. After allowing the reaction to proceed in the dark for 10 hours at room temperature, the diamond surface was rinsed with Phosphate Buffered Saline (PBS) for at least 3 times before either ESR, fluorescent, or Atomic Force Microscopy measurement.

\textbf{NV center-based ESR spectroscopy.}
The measurements were carried out on a home-built system following previously reported procedures \cite{Du2015Science}. Briefly, a 532 nm green laser was used to illuminate the NV centers for initialization and readout. The external magnetic field was set at $B_0=809$ G. The microwave (MW) and Radio frequency (RF) irradiations were generated by an arbitrary waveform generator (Agilent M8190a), amplified (Mini-circuits ZHL-20W-13+ for MW, ZHL-16W-43+ for RF) and delivered by a coplanar waveguide fabricated on a glass substrate. Double electron-electron resonance pulse sequences were used to detect the weak signal from the spin labels (\textbf{Fig. \ref{Fig:setup}d}). The MW pulses, which manipulate the NV-center for detection, was set at approximately 0.6 GHz. The RF pulse, which manipulates the spin label, was scanned between $2.15-2.39$ GHz. The $\pi-$pulses width of the MW and RF pulses was approximately 11 ns and 50 ns respectively, corresponding to Rabi frequencies around 45 MHz and 10 MHz, respectively. The phase accumulating duration $\tau_0$ between the MW $\pi/2-$pulses was 4 us (see \textbf{Supplementary Note 3A} for more details). The detected spectral signals were normalized by the amplitude of Rabi oscillations.

\textbf{Data availability.}
Data supporting the findings of this study are available within the article and its Supplementary Information file, and from the corresponding authors upon reasonable request.

\end{document}


\begin{titlepage}

\begin{center}

\textbf{\large Supplementary Information}
\begin{spacing}{2.0}

\end{spacing}

\textbf{\large Single DNA Electron Spin Resonance Spectroscopy in Aqueous Solutions}
\begin{spacing}{2.0}

\end{spacing}

Fazhan Shi,$^{1,2,3,*}$  Fei Kong,$^{1,2,3,*}$  Pengju Zhao,$^{1,3,*}$  Xiaojun Zhang,$^{4}$ Ming Chen,$^{1,3}$ Sanyou Chen,$^{1,3}$ \\Qi Zhang,$^{1,3}$ Mengqi Wang,$^{1,3}$ Xiangyu Ye,$^{1,2,3}$ Zhecheng Wang,$^{1,3}$ Zhuoyang Qin,$^{1,3}$ Xing Rong,$^{1,2,3}$ \\Jihu Su,$^{1,2,3}$ Pengfei Wang,$^{1,2,3}$ Peter Z. Qin,$^{4,\dag}$ and Jiangfeng Du,$^{1,2,3,\ddag}$

\begin{spacing}{1.5}
\end{spacing}

$^{1}$\emph {CAS Key Laboratory of Microscale Magnetic Resonance and Department of Modern Physics,\\
University of Science and Technology of China (USTC), Hefei 230026, China}\\
$^{2}$\emph {Hefei National Laboratory for Physical Sciences at the Microscale, USTC}\\
$^{3}$\emph {Synergetic Innovation Center of Quantum Information and Quantum Physics, USTC}\\
$^{4}$\emph {Department of Chemistry, University of Southern California, Los Angeles, CA 90089, USA}

\end{center}

\thispagestyle{empty}
\end{titlepage}

\tableofcontents

\setcounter{page}{1}
\newpage

\section{Characterizations of Diamond Nanopillars}

\begin{figure}[http]
\includegraphics[width=0.5\columnwidth]{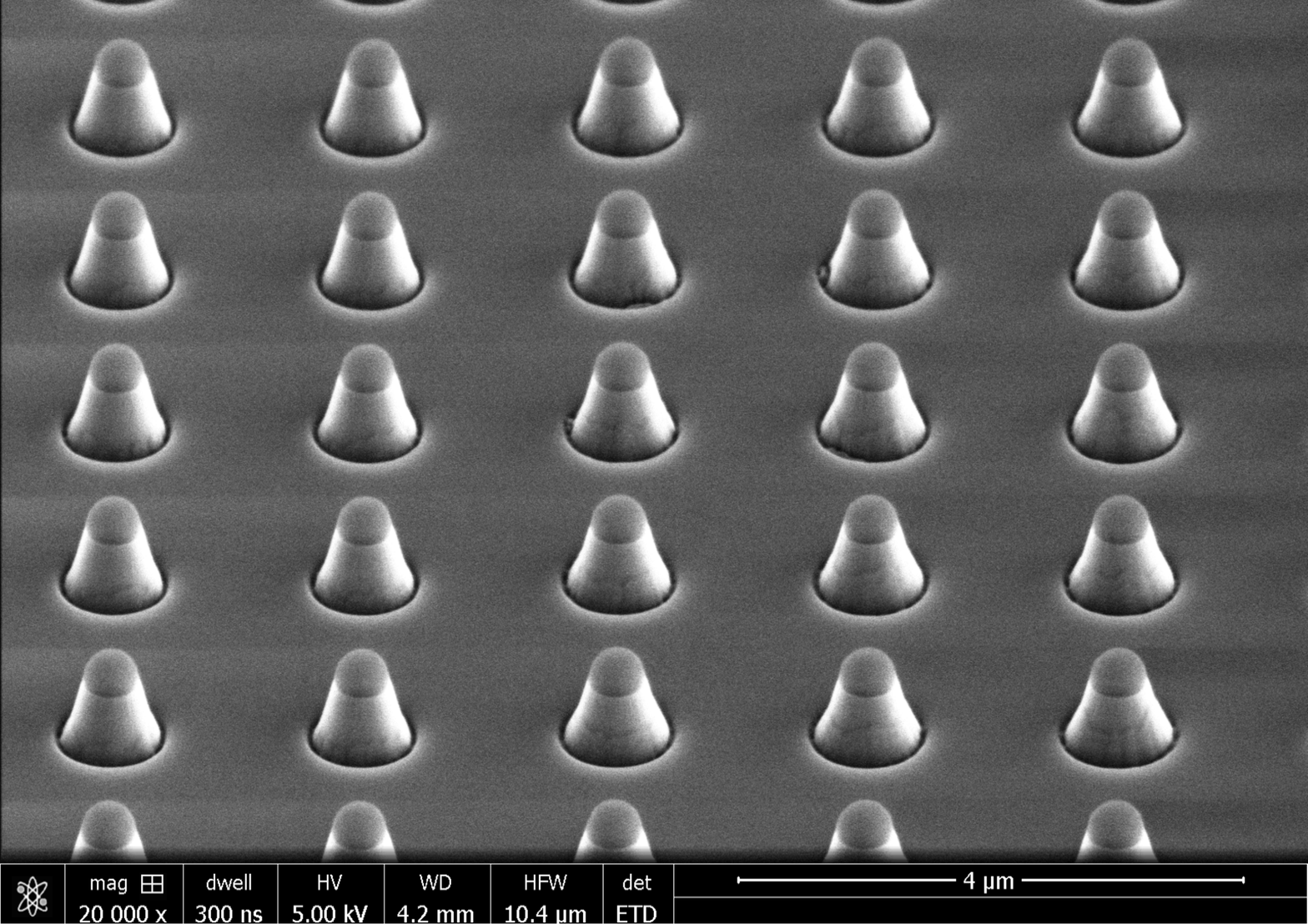}
\caption{\textbf{An example of an electron microscope image of the diamond nanopillars.}
}
\label{ElectromMicroscope}
\end{figure}

In this study, the pillars were fabricated so that the average pair-wise distance was approximate 2 micrometers (Supplementary Fig.~\ref{ElectromMicroscope}), and  approximately 50\% of them contained a single NV center. Given the confocal microscope used in the optically detected magnetic resonance has a 300 nm spatial resolution, pillars with NV sensors were well separated and can be examined one by one to detect the target spins, in this case, individual spin-labeled DNA duplexes.

\clearpage

\section{Characterizations of DNA Attachment at Diamond Surfaces}

\subsection{Surface topography characterized by Atomic Force Microscopy}

Atomic force microscopy (AFM) was carried out using a Dimension Icon setup (Bruker Corporation). An ATEC-NC probe (NANOSENSORS, Inc.) was used, which has a tip radius $<$ 10 nm and a resonance frequency of 335 KHz. The AFM images were acquired in the tapping mode. The diameter of DNA duplex is $\sim 2$ nm, which is smaller than the horizontal resolution of the AFM. As shown in Supplementary Fig.~\ref{AFM}a, individual DNAs can not be resolved on the diamond surface. In order to confirm the presence of the DNA, the surface was scraped with the AFM tip using the contacting mode. Supplementary Fig.~\ref{AFM}b shows an AFM image after scraping, which shows aggregations at the edge of the scraped area, reflecting the existence of the DNAs. In addition, the average height difference of the areas with and without scraping is $0.74$ nm, which is smaller than the expected width of a DNA duplex, suggesting the presence of a single layer of DNA at the diamond surface.

\begin{figure}[http]
\centering
\includegraphics[width=0.8\columnwidth]{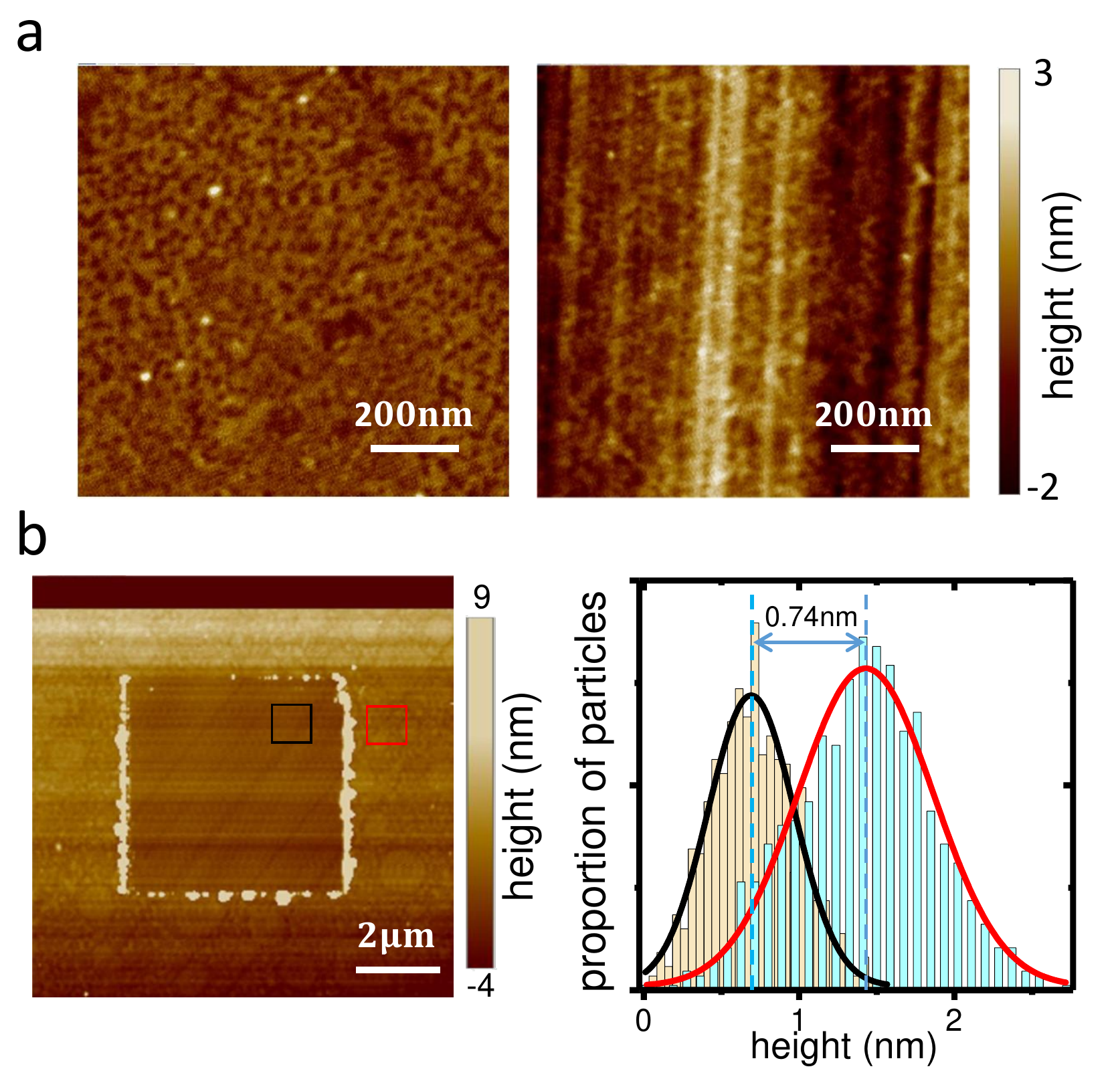} \protect\caption{\textbf{Diamond surface topography imaged by AFM.}
(\textbf{a}) Fine scanning of the surface of two diamonds after DNA duplex attachment.
(\textbf{b}) Image of a larger area after scrapping with the AFM tip. The height distributions inside and outside the scrapped area are shown on the right.
}
\label{AFM}
\end{figure}

\subsection{Surface density of the DNA estimated by confocal microscopy}

\begin{figure}[H]
\includegraphics[width=1\columnwidth]{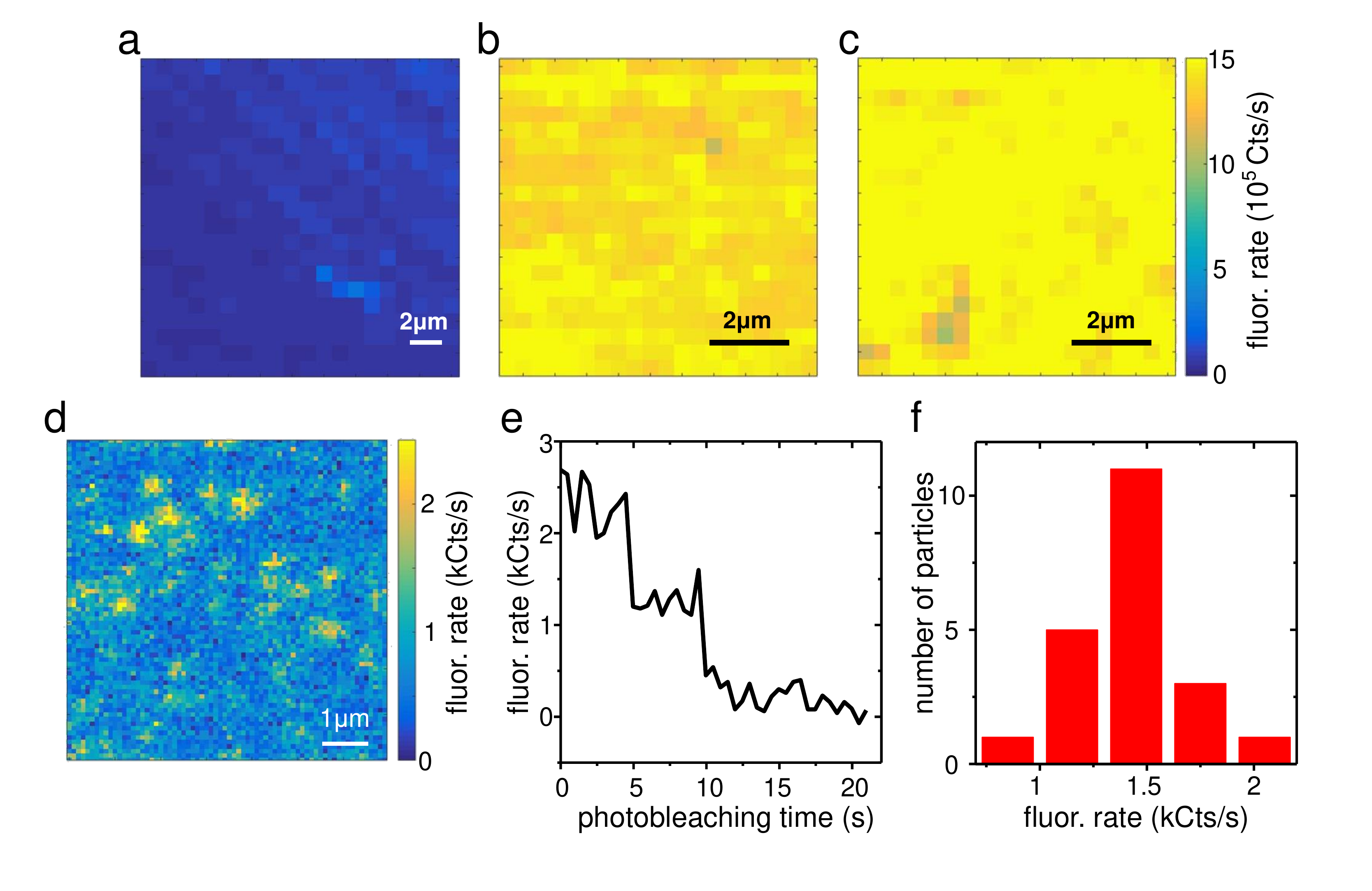} \protect\caption{\textbf{Fluorescence imaging of Cy3-s1 DNAs captured on diamond surfaces.}
(\textbf{a}) Confocal imaging of a surface without the tethered NH$_2-$DNA strand.
(\textbf{b})(\textbf{c}) Confocal imaging of two diamond surfaces with the tethered NH$_2$-DNA strand, which shows high fluorescence indicating capturing of the Cy3-s1 DNA strand.
(\textbf{d}) Confocal imaging of the diamond surface attached with diluted Cy3-s1 strand where single spots were observed.
(\textbf{e}) An example of the time trace of fluorescence at a single spot.
(\textbf{f}) Statistics of the fluorescence rate from spots showing a one-step jump.
}
\label{Cy3}
\end{figure}

Cy3-labeled DNA (Cy3-s1) was used to estimate the density of DNA on the diamond surface. Cy3-s1 was added onto different diamond surfaces, then rerinsed repeatedly with 100 mM NaCl solution. Fluorescence was measured using a home-build confocal microscope. The Cy3 dye was excited by a 2 $\mu $W, $532$ nm laser. The emitted photons were collected via an optical path that was divided into two by dichroic mirrors, with one path using a $640$ nm longpass filter to collect the fluorescence of the NV center and the other path using a $535-607$ nm bandpass filter to collect the fluorescence of Cy3. The data showed that in the absence of a tethered DNA strand (i.e., the `NH$_2$-DNA'), a low fluorescence rate of ~100 kCts/s was measured (Supplementary Fig.~\ref{Cy3}a), indicating that Cy3-s1 DNA was hardly captured at the surface. With a tethered NH$_2$-DNA strand, the fluorescence rate was measured to be ~1.5 MCts/s (Supplementary Fig.~\ref{Cy3}b, c), indicating efficient capture of the Cy3-s1 DNA. Images obtained from multiple diamonds indicated there was no observable difference between the different preparations (Supplementary Fig.~\ref{Cy3}b, c).

The fluorescence rate of individual Cy3-s1 DNA was detected by diluting the Cy3-s1 DNA with unlabeled DNA (Cy3 labeled DNA $:$ unlabeled DNA $=1:1000$). This resulted in observations of well-resolved spots (Supplementary Fig.~\ref{Cy3}d), although there also existed some areas with concentrated spots (data not shown). When tracing these spots, discrete jump in fluorescence were observed (Supplementary Fig.~\ref{Cy3}e) with the number of jump steps reporting the quantity of Cy3-s1 DNA at a given spot, while the height of the steps reflecting the fluorescence rate of single Cy3-s1 DNA, which was found to be $1 \sim 2$ kCts/s.
Furthermore, well-resolved spots showing a one-step jump were fit to a Gaussian surface. Using a histogram constructed from analyzing 21 spots showing a one-step jump (Supplementary Fig.~\ref{Cy3}f), the mean fluorescence rate of an individual Cy3-s1 DNA was determined to be $1.4\pm0.3$ kCts/s. Given that without dilution the fluorescence rate was measured to be 1.5 MCts/s, the number of DNA within the diffusion limited spot was calculated to be $1070 \pm 230$. With the diameter of the diffusion limited spot being $1.22 \lambda/2N.A. \approx 386$ nm (in this work fluorescence wavelength $\lambda \sim 570$ nm and numerical aperture of the lens $N.A. = 0.9$), the mean spacing between the DNA $a_0$ was calculated to be $21 \pm 2$ nm.

\clearpage

\section{Additional Data and Discussions on NV-Detected Nitroxide-Labeled DNAs}

\subsection{Additional discussions on NV-based detection of external nitroxide spin labels}

\begin{figure}[http]
\includegraphics[width=0.8\columnwidth]{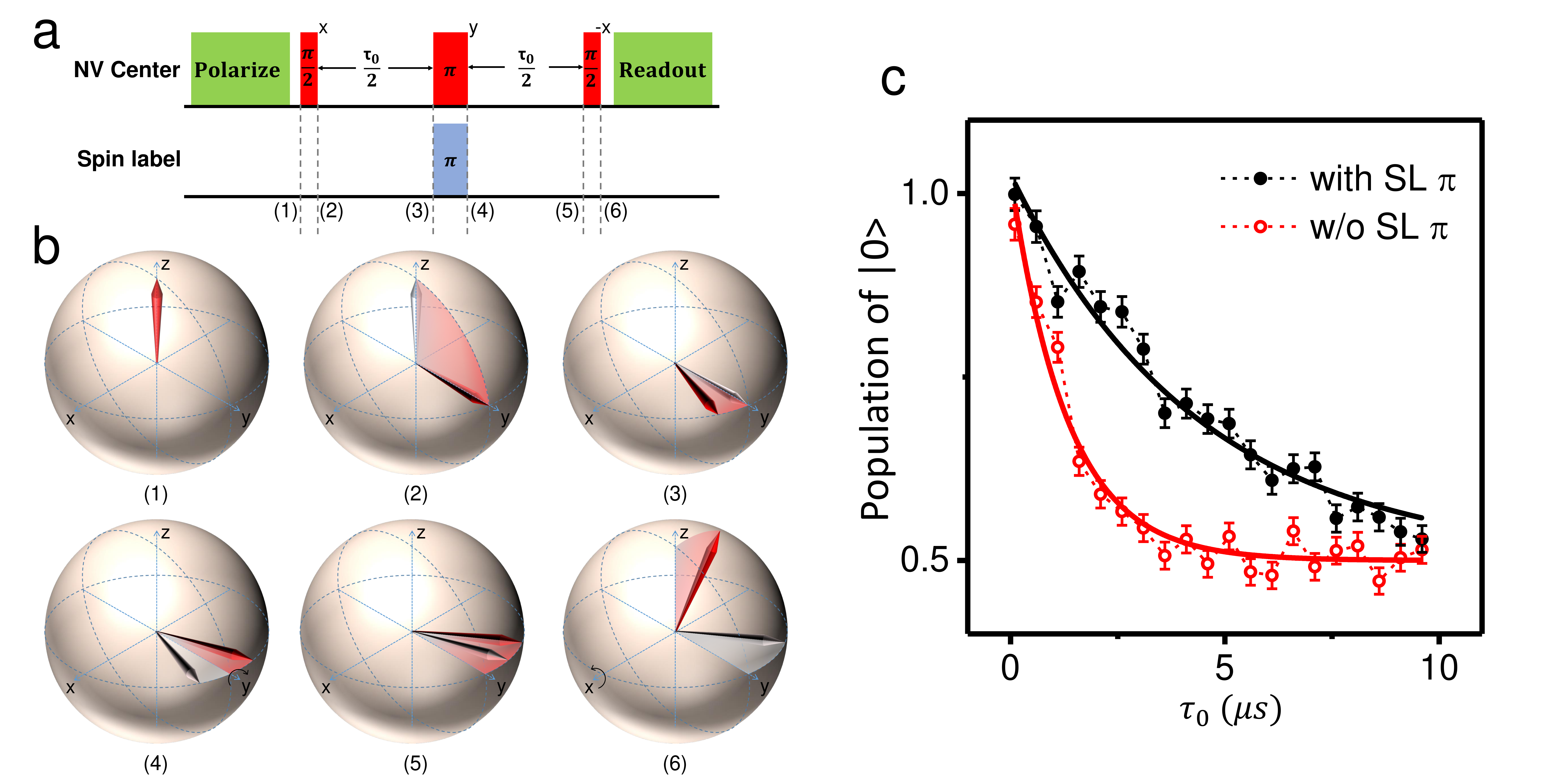} \protect\caption{\textbf{NV center-based detection method.}
(\textbf{a}) The double electron-electron resonance (DEER) pulse sequence.
(\textbf{b}) Corresponding evolution of the NV center spin represented on the Bloch sphere.
(\textbf{c}) Examples of observed echo evolution data. Without the $\pi$-pulse applied on the spin label (black dots and line), the measured decay trace reflected only the dechonerence of the NV center. With a $\pi$-pulse applied to successfully flip a spin label that was interacting with the NV (red dots and line), the NV center acquired an extra phase, resulting in a different decay. Error bars indicate $\pm$1 standard error of the mean (s.e.m) and the data shown were obtained from measurements repeated 1 million times.
}
\label{Principle}
\end{figure}

NV centers are color centers in diamonds consisting of a substitutional nitrogen atom and an adjacent vacancy. The electrons surrounding the defect form an effective electron spin with a spin triplet ground state ($S=1$), which can be polarized to $m_{\text{S}}=0$ with laser irradiation. Two of the ground states (i.e. $m_{\text{S}}=0$ and $m_{\text{S}}=-1$) can be treated as a pseudo spin half system.
The NV center functions as an interferometer (Supplementary Fig. \ref{Principle}).
The sensor spin (NV center, Supplementary Fig. \ref{Principle}a) was first polarized to $|0\rangle$ with a $1~\mu$s $100 ~\mu$W laser pulse, then rotated to $(|0\rangle+|-1\rangle)/\sqrt{2}$ with a microwave (MW) $\pi/2$ pulse. This superposition state would accumulate phase and become $(|0\rangle+e^{-i\phi}|-1\rangle)/\sqrt{2}$ with $\phi = \delta \phi + D\tau$, here $\delta \phi$ is the phase noise induced by bath spins and $D$ is the dipolar coupling between the NV center and the target spin. The central MW $\pi$ pulse was then applied to flip the superposition state, resulting in formation of an echo at the appropriate time. During the echo evolution, the phase noise was cancelled to a large degree, while the coupling term $D$ would survive only if the radiofrequency (RF) pulse would flip the target spin (e.g. "Spin label", Supplementary Fig. \ref{Principle}a). Finally, the superposition state was rotated back to the initial state with a MW $\pi/2$ pulse (Supplementary Fig. \ref{Principle}a), and the population of $|0\rangle$ can be measured by fluorescence. After repeating the above sequence for several million times, signal appears only when the frequency of the RF pulse is on resonance with transitions of the target spin. By scanning the frequency of RF, a resonance spectrum was obtained (See Main text Fig. 2 and Supplementary Note 3C).

In this work, to optimize detection of the target spin ("Spin label") spectrum, echo evolutions were first measured by fixing the RF frequency at the center peak transition of the $^{14}$N nitroxide while increasing step by step the evolution time $\tau_0$ between the MW $\pi/2$ pulses (Supplementary Fig. \ref{Principle}a). This was equivalent of moving the RF pulse on-resonance with the nitroxide spin label. Supplementary Fig. \ref{Principle}c shows examples of echo evolution data obtained, with the maximal difference with and without the RF pulse applied observed at $\tau_0=4~\mu$s. Therefore, in subsequent spectral measurements, $\tau_0$ was fixed at 4 $\mu$s while the RF frequency was scanned.

\clearpage

\subsection{Statistical probability of detecting a single DNA}

For the pulse sequence (Supplementary Fig.~\ref{Principle}a) used in this work, the acquired signal contrast was
\begin{equation}
S = P(\frac{1-cos\phi_0}{2})e^{-(\tau_0/T_2)^{p}},
\label{signal}
\end{equation}
here $P=\langle\uparrow|U_{\text{rf}}|\downarrow\rangle$ is the flipped probability of a target spin manipulated by radiofrequency, $\tau_0$ is the evolution time, $\phi_0 = D\tau_0$ is the accumulated phase of the NV center spin during the evolution, which is proportional to the dipolar coupling $D$ between the NV center and the target spin, $T_2$ is the coherence time of the NV center, the stretched exponential $e^{-(\tau_0/T_2)^{p}}$ ($p$ in the range of $1-3$) is determined by the dynamic of bath \cite{Walsworth2012NatCommu}.
For the NV centers used in this work, $p\sim 1$.The mean coherence time and evolution time are $T_{2}\sim 4$ $\mu$s and $\tau_0\sim 4$ $\mu$s, respectively. The dipolar coupling is given by
\begin{equation}
D = \frac{D_0}{r^3}|1-3\cos^{2}\theta|,
\label{dipolar}
\end{equation}
where $D_0 = 52$ $\text{MHz}\cdot\text{nm}^{3}$, $r$ is the distance between the target spin and the NV center, $\theta$ is the angle between $\mathbf{r}$ and the magnetic field. Supposing the NV center and the target spin are located at $(0,0,-\Delta d)$ and $(X,Y,0)$, as shown in Supplementary Fig.~\ref{Signal_area}a, the dipolar coupling can be calculated as
\begin{equation}
D(X,Y) = D_0\frac{|2XY+2X\Delta d+2Y\Delta d|}{(X^2+Y^2+\Delta d^2)^{\frac{5}{2}}},
\label{dipolar}
\end{equation}
where the direction of the magnetic field is set as $(1,1,1)/\sqrt{3}$ because the diamond is 100-orientated. The mean depth of the NV centers is $\sim 6$ nm due to an implantation energy of 2 keV per ion \cite{Lehtinen2016}, and the distance between the spin label and the attached point at the diamond surface is $\sim 4$ nm. After averaging over the half sphere, the mean height difference between the NV center and the spin label is $\Delta d \sim 6+4\times0.5 = 8$ nm. Supplementary Fig.~\ref{Signal_area}b and c give the calculated dipolar coupling strength and the corresponding signal contrast, respectively, which clearly shows  that the NV center can only detect the nearby spins. Specifically, a spin is detectable only when the signal induced by this spin is stronger than the photon shot noise. As shown in the next section, the noise level in this experiment is $\sim 0.05$. Considering the spectrum of $^{14}$N nitroxide has three peaks, the detectable signal induced by a single nitroxide need to be $> 0.15$. Therefore, the detection area $\mathscr{S}$ is define by $S(X,Y)|_{{X,Y}\in\mathscr{S}}>0.15$, resulting in an area of $\|\mathscr{S}\|=72$ nm$^{2}$ (Supplementary Fig.~\ref{Signal_area}c).

\begin{figure}[H]
\includegraphics[width=1\columnwidth]{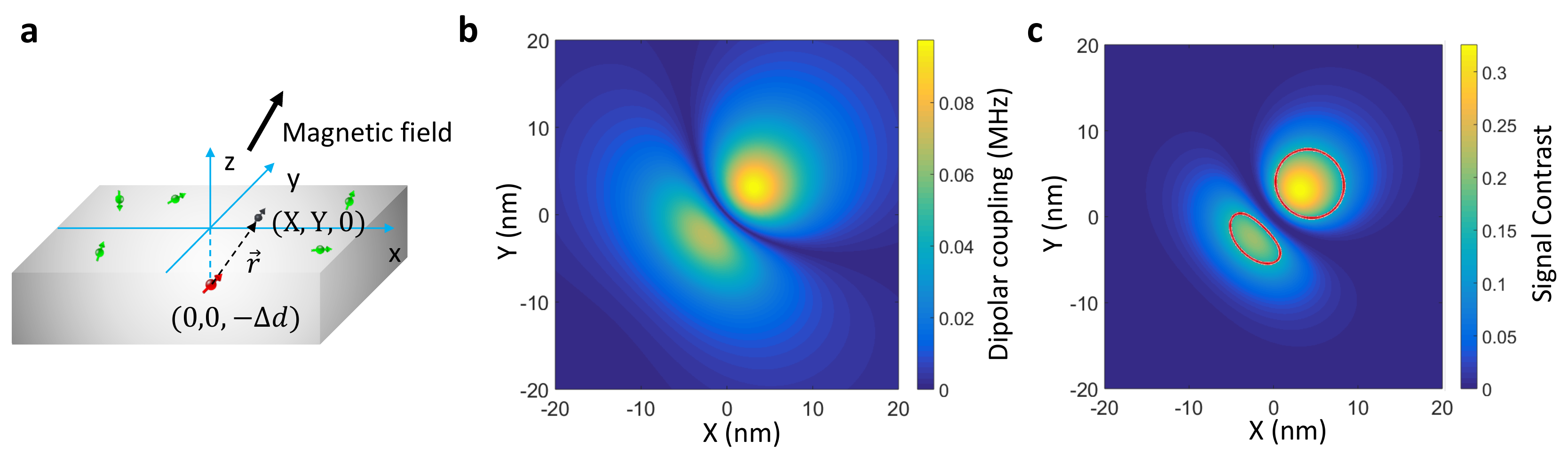} \protect\caption{\textbf{Calculation of the signal of electron spin dispersed on the diamond surface.}
(\textbf{a}) Schematic diagram of the position of the NV center (red arrow) and the surface spins (green arrows). The NV center is located at the center with the N-V axis lying along the 111 direction.
(\textbf{b}) Dipolar coupling at different positions.
(\textbf{c}) Corresponding signal constrast at different positions. The red circles mark the signal threshold of 0.15.
}
\label{Signal_area}
\end{figure}

Since the mean DNA spacing is $a_0 = 21$ nm (Supplementary Note 2B), corresponding to a DNA density of $\sigma = a_0^{-2}=2.27\times10^{-3}$ nm$^{-2}$, there exists a mean number of $N$ of DNAs in an area of $N/\sigma$. The probability of $k$ DNAs located in the detection area obeys the binomial distribution
\begin{equation}
p(k) = {N\choose k}(\frac{\|\mathscr{S}\|}{N/\sigma})^k(1-\frac{\|\mathscr{S}\|}{N/\sigma})^{N-k}.
\end{equation}
In the limit of $N\rightarrow \infty$, it approaches a Poisson distribution
\begin{equation}
p(k) = \frac{\eta^k e^{-\eta}}{k!},
\label{prob_of_k}
\end{equation}
where $\eta = \sigma \|\mathscr{S}\| = 0.16$. Therefore, the probability of one DNA located in the detection area is $p(1)=14\%$, and the probability of two or more DNAs located in the detection area is $1-p(0)-p(1)=1\%$.

If $N$ electron spins are detected, the signal given by Supplementary Eq.~\ref{signal} need to be modified to \cite{Wrachtrup2017SciAdv}
\begin{equation}
S_{\text{sum}} = P(\frac{1-\prod_{j}^{N}\cos{D_{j}\tau_0}}{2})e^{-(\tau_0/T_2)^{p}}.
\label{signal2}
\end{equation}
For the spins outside the detection area, $D_{j}\tau_0 \ll 1$, then Supplementary Eq.~\ref{signal2} can be simplified to
\begin{equation}
\begin{split}
S_{\text{sum}} &= P[\frac{1-\prod_{j}^{N}(1-\frac{D_{j}^{2}\tau_{0}^{2}}{2})}{2}]e^{-(\tau_0/T_2)^{p}}\\
&= P[\frac{1-(1-\sum_{j}^{N}\frac{D_{j}^{2}\tau_{0}^{2}}{2})}{2}]e^{-(\tau_0/T_2)^{p}}\\
&= \sum_{j}^{N}P\cdot\frac{D_{j}^{2}\tau_{0}^{2}}{4}\cdot e^{-(\tau_0/T_2)^{p}}=\sum_{j}^{N}S_{j}.
\end{split}
\label{signal3}
\end{equation}
So the signal contributed by the spins outside the detection area can be calculated by
\begin{equation}
S_{\text{out}} = \int_{{X,Y}\notin\mathscr{S}}S(X,Y)\sigma dX dY \approx 0.05,
\label{background}
\end{equation}
This indicates that all spins outside the detection area will induce a signal comparable to the photon shot noise ($\sim$0.05) and therefore is not detectable. Specifically, the signal of a single DNA is $>0.1\times3$ (see Supplementary Note 3C), so more than ($0.3/[0.3+0.05]=$) $86\%$ of the signal is attributable to the signal DNA.

\clearpage

\subsection{Statistics of the NV-detected spectra}

\begin{figure}[http]
\includegraphics[width=0.8\columnwidth]{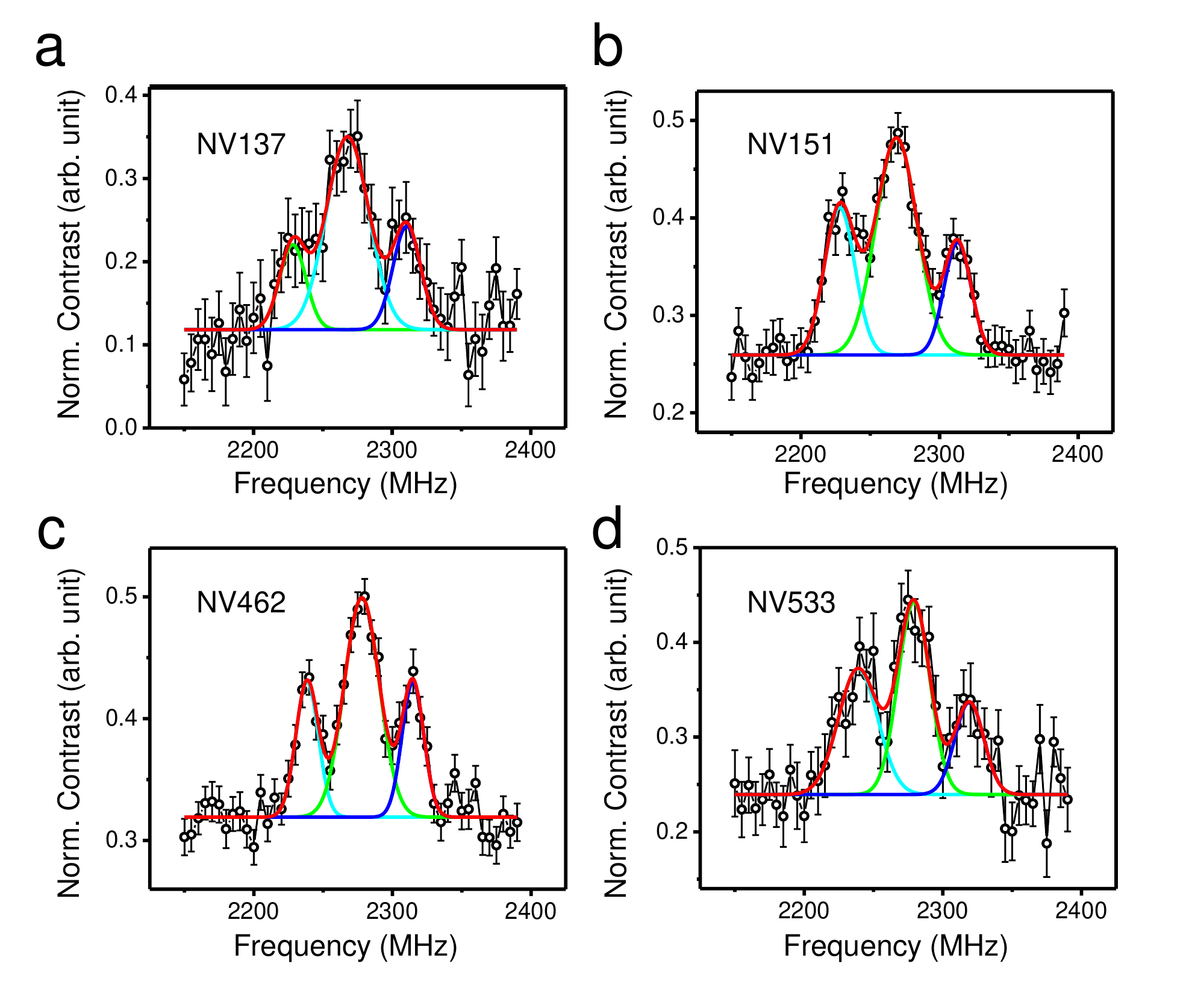} \protect\caption{\textbf{Additional NV-detected single $^{14}$N-R5 labeled DNA ESR spectra.}
The measured hyperfine splitting is (\textbf{a}) $41.1\pm3.0$ MHz, (\textbf{b}) $42.5\pm1.2$ MHz, (\textbf{c}) $38.1\pm1.0$ MHz and (\textbf{d}) $40.1\pm2.6$ MHz. The contrast is normalized by the the amplitude of Rabi oscillation and the different positions of the baseline indicate different coherence properties of different NV centers. Error bars indicate $\pm$1 s.e.m and the data shown were obtained from measurements repeated 2 million times.
}
\label{ESR_14N}
\end{figure}

Measurements of $^{14}$N-R5 labeled DNA were carried out on 97 NV centers, 12 of the NV centers exhibited side peaks in the first round of scan, which lasted for $\sim 20$ minutes. The proportion of NV centers showing DNA signal is similar to the probability of one DNA located within the detection area estimated in Supplementary Note 3B. After repeating the measurements several times to achieve a suitable  signal-to-noise ratio (SNR), only five NV-detected spectra exhibited clear side peaks (Main text Figure 2a and Supplementary Fig.~\ref{ESR_14N}). Hyperfine splitting among the NV-detected spectra varied, which may reflect different polarity of the environment at the individual DNAs (see Supplementary Note 5A).
In addition, ratios of the signal contrast between the individual peaks also varied. It is not clear whether this is a result of the limited SNR, or reflects variations among the individuate DNAs, such as the difference in positioning with respect to the NV center (see Supplementary Note 3E) or rotational dynamics (see Supplementary Note 5B). Further investigations are needed. The signal of $^{15}$N-R5 labeled DNA were also observed at multiple cases (see Supplementary Fig.~\ref{ESR_15N}).

\begin{figure}[http]
\includegraphics[width=0.8\columnwidth]{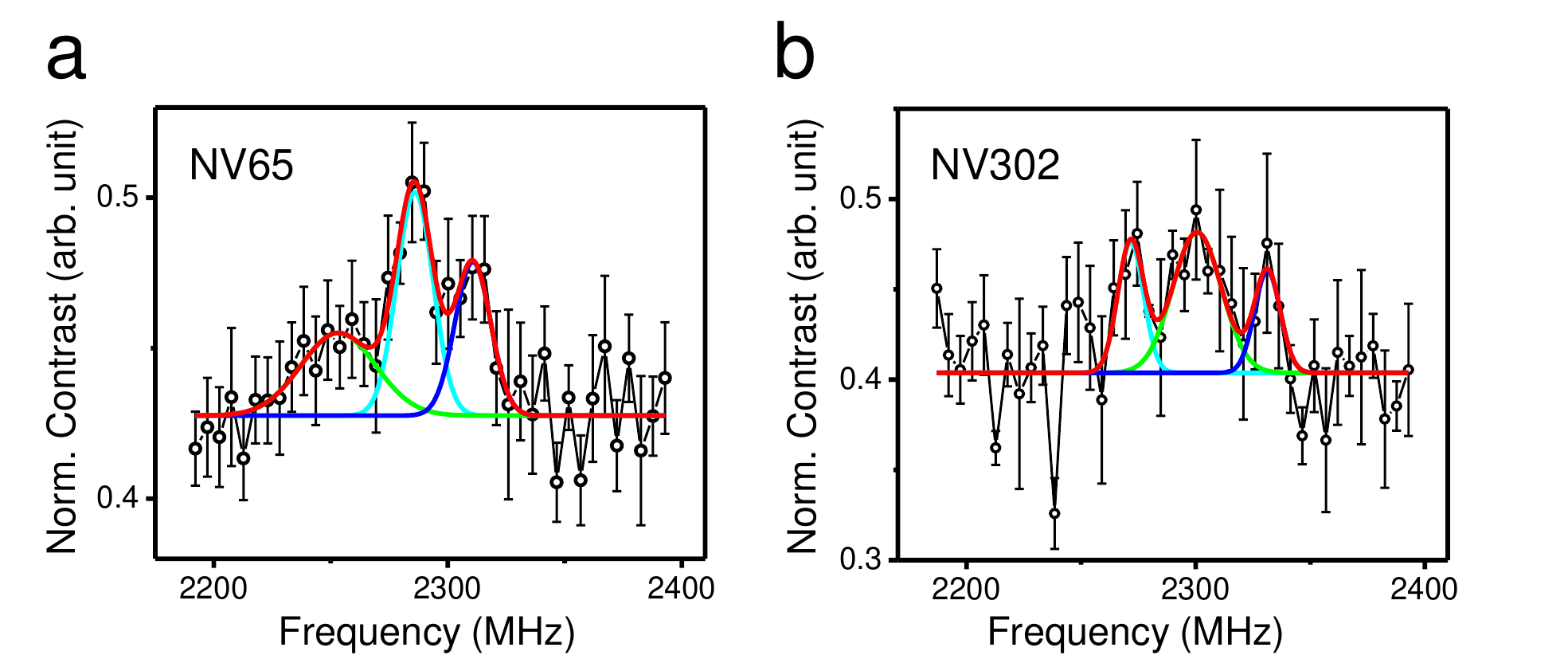} \protect\caption{\textbf{Additional NV-detected single $^{15}$N-R5 labeled DNA ESR spectra.}
The measured hyperfine splitting is (\textbf{a}) $57.4\pm8.9$ MHz and (\textbf{b}) $56.8\pm5.2$ MHz. The contrast is normalized by the the amplitude of Rabi oscillation. Error bars indicate $\pm$1 s.e.m and the data shown were obtained from measurements repeated 4 and 1.5 million times, respectively, for panel (a) and (b).
}
\label{ESR_15N}
\end{figure}
\clearpage

\subsection{Control measurements without the R5-labeled DNA strand}

As controls, we also measured the signal of "clean" diamond surfaces with only the NH$_{2}$-DNA tethered. Only a single peak was observed (see Supplementary Fig.~\ref{ESR_empty}a), which is similar to the cases observed after the spin labels were quenched by laser irradiation (see Supplementary Fig.~\ref{ESR_empty}b).

\begin{figure}[http]
\includegraphics[width=0.8\columnwidth]{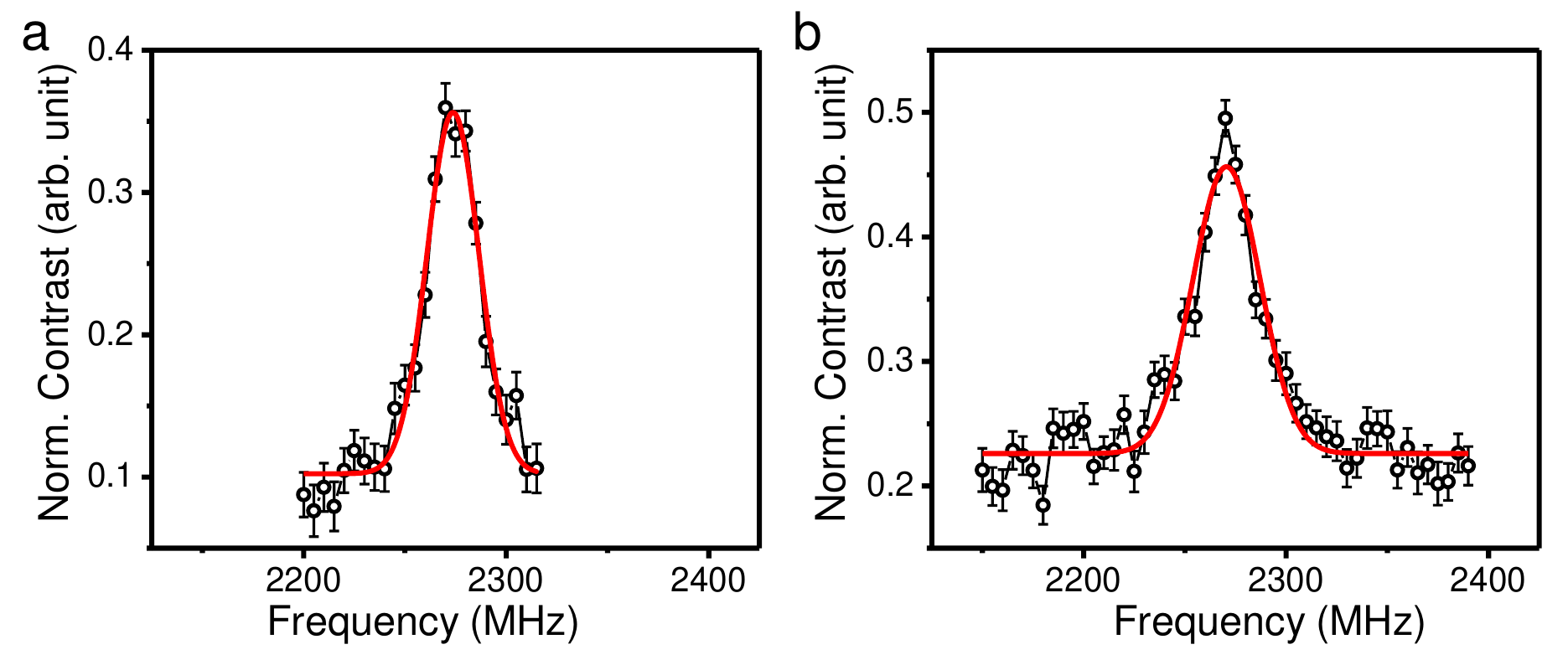} \protect\caption{\textbf{Background ESR signal.}
(\textbf{a}) Signal observed when only the NH$_{2}$-DNA was tethered on the diamond surface without hybridizing to a spin-labeled complementary strand.
(\textbf{b}) Signal observed after laser bleaching of a surface tethered with spin-labeled DNAs. The data shown were obtained from measurements repeated 2 million times. Error bars indicate $\pm$1 s.e.m.
}
\label{ESR_empty}
\end{figure}

\clearpage

\subsection{Additional discussions on signal intensity of the DNA and the surface defects}

The NV-detected DNA spectra (Main text Fig. 2 and Supplementary Note 3C) show variations in the normalized signal contrast between the side-peaks and the central peak. Such variations likely arise from two main factors: (1) the presence of the paramagnetic surface defects, which impact the center peak but not the side-peaks; and (2) variations in the positioning of the individual spin labels (or DNAs) with respect to the NV center.

As shown in Supplementary Eq. \ref{signal3}, the total signal of many spins is equal to the sum of the signal of each spin only when the signal induced by each spin is sufficiently small. However, for the shallow NV centers, the dipolar coupling with a target spin at the diamond surface can be approximately 0.1 MHz and the coherence time is roughly 4 $\mu$s. So the phase $D_{j}\tau_0$ is on the order of 1, and as shown in Supplementary Eq. \ref{signal} and Eq. \ref{signal2}, the signal can has negative correlation with the quantities of coupling spins when $D_{j}\tau_0 > \pi/2$. Supposing the mean spacing of background spins is 10 nm \cite{Degen2014PRL}, the signals of a single spin and an ensemble of spins are shown in Supplementary Fig.~\ref{signal_sim}. For a single spin, the maximum signal is 1 (if no decoherence). However, the maximum signal will reduce to 0.5 if many spins are involved. At the central line, many background spins and a single DNA contribute to the signal, so the maximal signal $< 0.5$. At the side line, only a single DNA contributes the signal, so the maximal signal $< 1$. Considering the three (two) nuclear spin states for 14N (15N), the maximal signal of the side line $< 0.33~(0.5)$. If the position of the spin label (i.e., the DNA) is close enough to the NV center, it is possible that the side line has a similar amplitude as that of central line. Also, all patterns of normalized signal contrast shown in Main text Fig. 2 and Supplementary Note 3C are possible if the position of the spin label is suitable.

\begin{figure}[http]
\includegraphics[width=0.8\columnwidth]{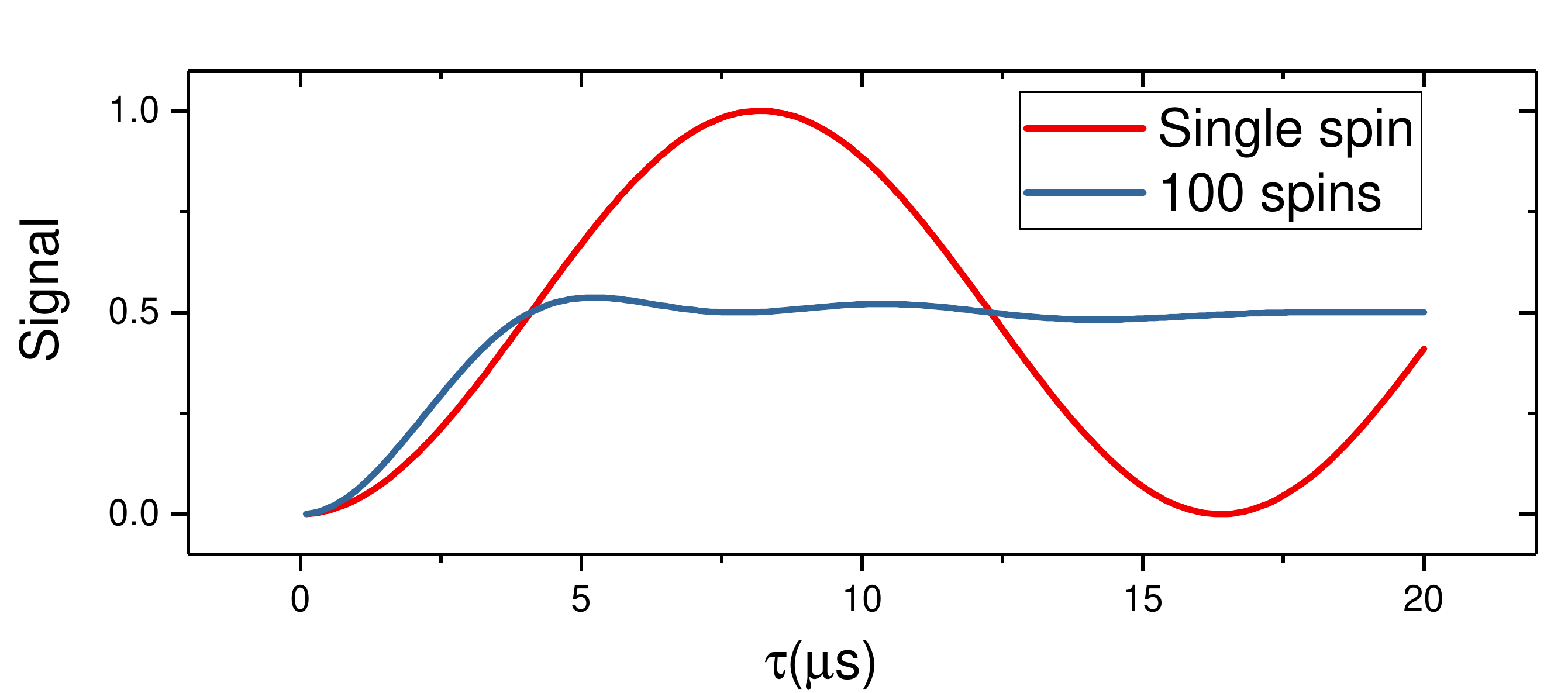} \protect\caption{\textbf{Simulated signals of a single spin (red) and an ensemble of electron spins (green).}
}
\label{signal_sim}
\end{figure}

\clearpage

\section{Ensemble X-Band Continuous-Wave ESR Spectroscopy}

For bulk ensemble ESR measurements, DNA duplexes were formed in a buffer solution of 0.1 M sodium phosphate (pH 7.0) following reported procedures \cite{Ding2014}, with the ratio of the spin-labeled DNA to its non-labeled complementary strand being approximately $1:1.5$. X-band continuous-wave (cw) EPR spectra were measured at room temperature on a Bruker EMX spectrometer equipped with an ER4119HS resonator. The incident microwave power was 2 mW, and the field modulation was 1 G at a frequency of 100 kHz. Each spectrum was acquired with 512 points, corresponding to a spectral range of 100 G. All spectra were corrected for background and baseline, then normalized to the same number of spins following reported procedures \cite{Zhang2009}.

\begin{figure}[http]
\includegraphics[width=0.6\columnwidth]{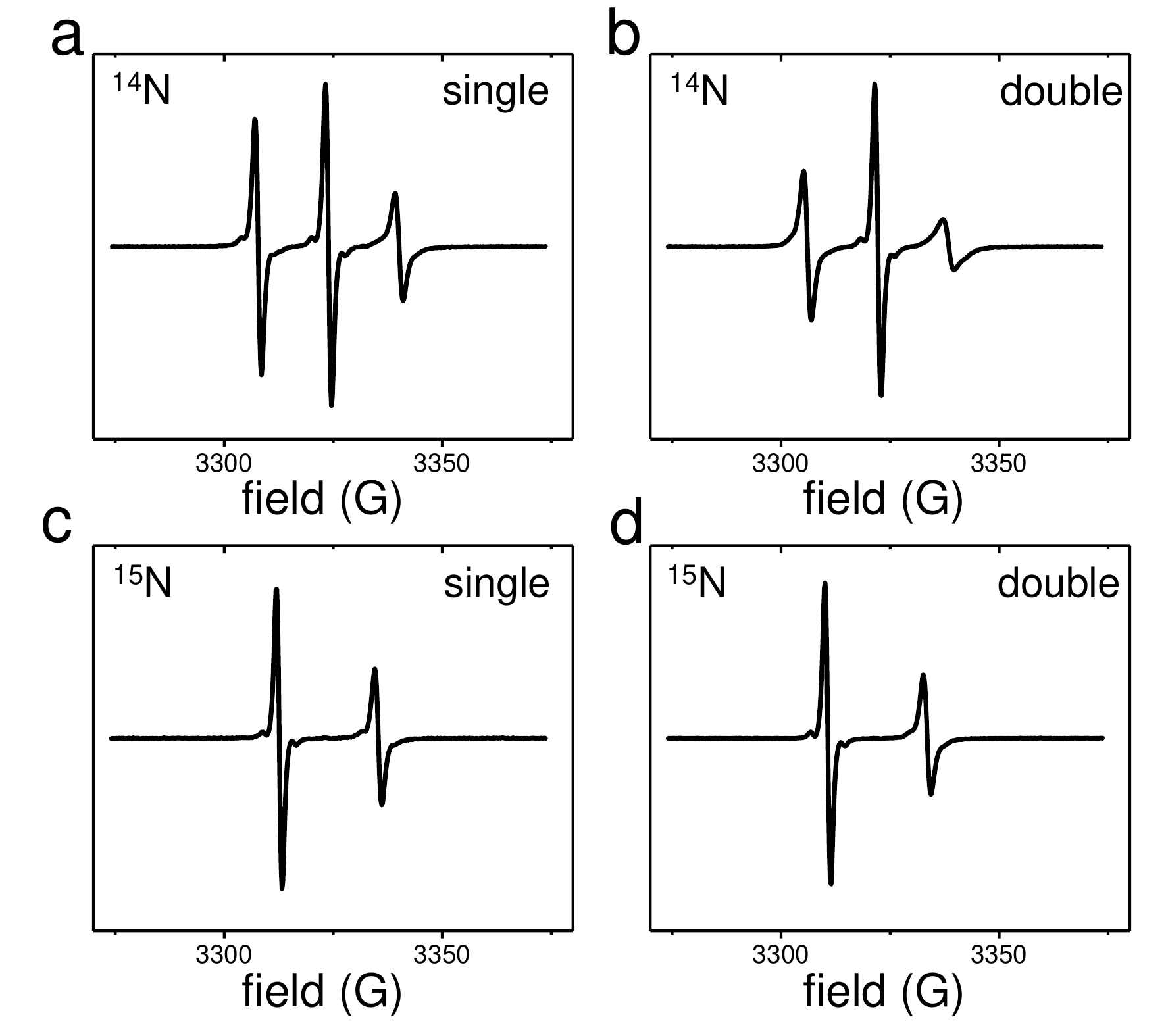}
\caption{\textbf{Bulk X-band cw-EPR spectra of DNAs labeled with $^{14}$N-R5 or $^{15}$N-R5.}
These spectra were obtained with $50-100 ~\mu$M DNA in aqueous buffer. The spectra are shown in the conventional derivative representation. Corresponding first-integral spectra (i.e., the absorption spectrum) of the double-stranded DNAs are shown in Fig.~2(b)(d) in the main text.
}
\label{Ensemble_ESR}
\end{figure}

Supplementary Fig.~\ref{Ensemble_ESR} shows the spectra of single-stranded and double-stranded DNAs labeled with R5. As expected, the $^{14}$N ($I=1$) spectra show a three-line splitting (panel a, b), while the $^{15}$N ($I=1/2$) spectra show a two-line splitting (panel c, d). The spectra show un-even sharp lines, which are consistent with previous results \cite{Popova2009}. Spectra of the double-stranded DNAs (panel b, d) show broader lines as compared to that of the corresponding single-stranded DNAs (panel a, c), consistent with a reduction in rotational motion of the DNA upon duplex formation.

\clearpage

\section{Dynamic Information Derived from NV-Detected ESR Spectrum}

\subsection{Qualitative assessments of spectral lineshape}

The hyperfine tensor of the nitroxide spin label is anisotropic. If the nitroxide is stationary or undergoing slow motions, one would expect that for differently orientated spin labels, the NV-detected spectra would show large differences in the spacing between the peaks, as well as variations in lineshape of the side-peaks. This was indeed observed in our previous work on spin-labeled proteins fixed within a poly-Lysine layer \cite{Du2015Science}. However, in current work, all side-peaks in the NV-detected spectra shown a single-peak lineshape (Main test Fig. 2 and Supplementary Note 3C), although small variations can be detected in the linewidths and spacing between the side-peaks. The pattern observed in the NV-detected spectra is characteristically identical to those previously observed in bulk measurements of R5-labeled DNA duplexes in solutions \cite{Popova2009} and tethered to nanodiamonds \cite{DNA_nanodiamond}. It has been established that the line-pattern indicates that the nitroxide label is undergoing fast isotropic rotational motions, which result in nearly complete average of the anisotropic hyperfine tensor to an isotropic $A_{\text{iso}}$ value \cite{Popova2009}\cite{DNA_nanodiamond}. As already discussed in the nanodiamond work \cite{DNA_nanodiamond}, even the DNA is tethered, the flexibility of the chemical linker and the motions of the nitroxide ring with respect to the DNA allow the nitroxide to sample nearly all possible orientations with respect to the external magnetic field in a fast rate. This accounts for the charactersitc lineshape of the NV-detected spectra of diamond-tethered DNAs reported in this work (see Supplementary Note 5B below).

We do note that the NV-detected single-DNA spectra show slight variations in hyperfine splitting $A_{\text{iso}}$ (Supplementary Fig.~\ref{ESR_14N} and Supplementary Fig.~\ref{ESR_15N}), reflecting heterogeneity among the individual molecules in solutions. It is possible that the $A_{\text{iso}}$ variations are due to differences in the motions between individual molecules. However, one would then expect to observe some cases where the $A_{\text{iso}}$ values are greater than that measured in the bulk spectrum, which represents the ensemble average of the individual $A_{\text{iso}}$. This was not observed, but instead, all the $A_{\text{iso}}$ values from the single DNA spectra were smaller than that measured in the bulk: for the $^{14}$N-R5 labeled DNA, $A_{\text{iso}}$ varied between 38.1 to 42.5 MHz (Supplementary Fig.~\ref{ESR_14N}), while the bulk value was 45.2 MHz (Supplementary Fig. \ref{Ensemble_ESR}b); and similar behaviors were observed for the $^{15}$N-R5 labeled DNAs (Supplementary Fig.~\ref{ESR_15N} and \ref{Ensemble_ESR}d). Therefore, the data indicated that other factors, such as local variations of hydrophobicity (see main text), are more likely to account for the observed $A_{\text{iso}}$ variations.

Another difference between the NV-detected and bulk solution spectra is line-broadening of the side-peaks. For the NV-detected $^{14}$N-R5 spectrum shown in Fig.~2a in the main text, the linewidth (full width at half maximum, FWHM) of left, center and right peaks are $20.7\pm2.9$, $23.5\pm2.5$ and $16.7\pm2.1$ MHz, respectively. while the corresponding values for the bulk ensemble spectrum are $7.5\pm0.1$, $5.0\pm0.1$ and $13.5\pm0.2$ MHz, respectively. In the NV experiments, the nitroxide electron spin was flipped by a RF pulse (Supplementary Fig. \ref{Principle}a), and the peak width is influenced by the RF power. All NV-detected spectra reported in this work were acquired with a 50 ns RF pulse, corresponding to a RF-induced line broadening of 10 MHz line. The approximately 20 MHz peak width observed in the NV-detected spectra (Fig.~2a in the main text) is larger than the RF-induced broadening, and likely reflects retarded rotational motions of spin label when attached to the surface tethered DNA duplexes.

\subsection{Simulation of NV-detected spectra}

Simulations based on an isotropic rotation model are presented here. The use of a simple isotropic rotation model is appropriate based on qualitative lineshape analyses (Supplementary note 5A). The spin label used in this work is a nitroxide radical with a unpaired electron spin ($S=1/2$) and a nitrogen nuclear spin ($I=1$ for $^{14}$N). In the lab frame with a magnetic field along the Z axis, the Hamiltonian of the spin label is
\begin{equation}
H_{\text{SL}}=\mathbf{B}\cdot\mathbf{g}\cdot\mathbf{S}+\mathbf{S}\cdot\mathbf{A}\cdot\mathbf{I}+{\gamma_{\text{n}}}{B_{0}}{I_{z}}+\mathbf{I}\cdot\mathbf{Q}\cdot\mathbf{I}.
\end{equation}
Here $\mathbf{B}=(0,0,B_0)$ is the magnetic field, $\mathbf{S}$ and $\mathbf{I}$ are electron and nuclear spin operators, $\gamma_{\text{n}}$ is the gyromagnetic ratio of the nuclear spin, $\mathbf{g}$ is the Lande factor, $\mathbf{A}$ is the hyperfine tensor, and $\mathbf{Q}$ is the $^{14}$N nuclear quadrupole tensor. The last two terms does not contribute to the allowed electron resonance lines ($\Delta I=0$). The hyperfine tensor can be divided into two parts
\begin{equation}
\mathbf{A}=A_{\text{iso}}\cdot \mathbf{I}+|T_{d}|\cdot \mathbf{\Gamma}_{d},
\end{equation}
where $A_{\text{iso}}$ is the isotropic hyperfine coupling, $\mathbf{I}$ is the identity matrix, $|T_{d}|$ is the strength of the electron-nucleus dipolar interaction and $\mathbf{\Gamma}_{d}$ is the traceless hyperfine tensor. In the principle axis frame of the nitroxide, both $\mathbf{g}$ and $\mathbf{\Gamma}_{d}$ are diagonal
\begin{equation}
\mathbf{g}^{\text{PA}}=
\left(\begin{array}{ccc}
g_{xx} &  & \\
 & g_{yy} & \\
 &  & g_{zz}
\end{array}\right),
\mathbf{\Gamma}_{d}^{\text{PA}}=
\left(\begin{array}{ccc}
-1 &  & \\
 & -1 & \\
 &  & 2
\end{array}\right).
\end{equation}
For the R5 spin-label, $g_{xx} \approx 2.0083$, $g_{yy} \approx 2.0051$, $g_{zz} \approx 2.0022$, $A_{\text{iso}} \approx 44.5$ MHz, $|T_{d}| \approx 26.8$ MHz \cite{Qin2010}, all of these values are slightly different in different polarity environment \cite{Kurad2003}. For example, when simulating the spectrum shown in Fig. 2a in the main
text, $A_{\text{iso}} = 38.4$ MHz and $|T_{d}| = 23.1$ MHz.
In the lab frame, $\mathbf{g}=\mathbf{R}\cdot\mathbf{g}^{\text{PA}}\cdot\mathbf{R}^{-1}$,
$\mathbf{\Gamma}_{d}=\mathbf{R}\cdot\mathbf{\Gamma}_{d}^{\text{PA}}\cdot\mathbf{R}^{-1}$, where $\mathbf{R}$ is the rotation matrix
\begin{equation}
\mathbf{R}=
\left(\begin{array}{ccc}
\cos{\theta}\cos{\phi} & -\sin{\phi} & \sin{\theta}\cos{\phi}\\
\cos{\theta}\sin{\phi} & \cos{\phi} & \sin{\theta}\sin{\phi}\\
-\sin{\theta} & 0 & \cos{\theta}
\end{array}\right)
\end{equation}
determined by the direction of the principle axis ($\theta$, $\phi$). Rotation of the spin vector includes the tumbling of the DNA and the motion of R5 with respective to the DNA, which can be described by a set of random $\{\theta(t),~\phi(t)\}$ varying with time $t$ \cite{MonteCarlo}. The variation speed is described by the correlation time $\tau$ of the random walk. As described above, the signal measured by NV center is proportional to the flip probability of the labeled spin $P=\langle\uparrow|U_{\text{rf}}|\downarrow\rangle$, with
\begin{equation}
U_{\text{rf}}=e^{-i\int\{H_{\text{SL}}[\theta(t),\phi(t)]+H_{1}(t)\}dt},
\end{equation}
here $H_{1}(t)=\gamma_{\text{e}}B_{1}\cos{\omega t}S_{x}$ is the interaction between nitroxide electron spin and RF field with amplitude $B_{1}$ and frequency $\omega$.

\begin{figure}[H]
\includegraphics[width=1\columnwidth]{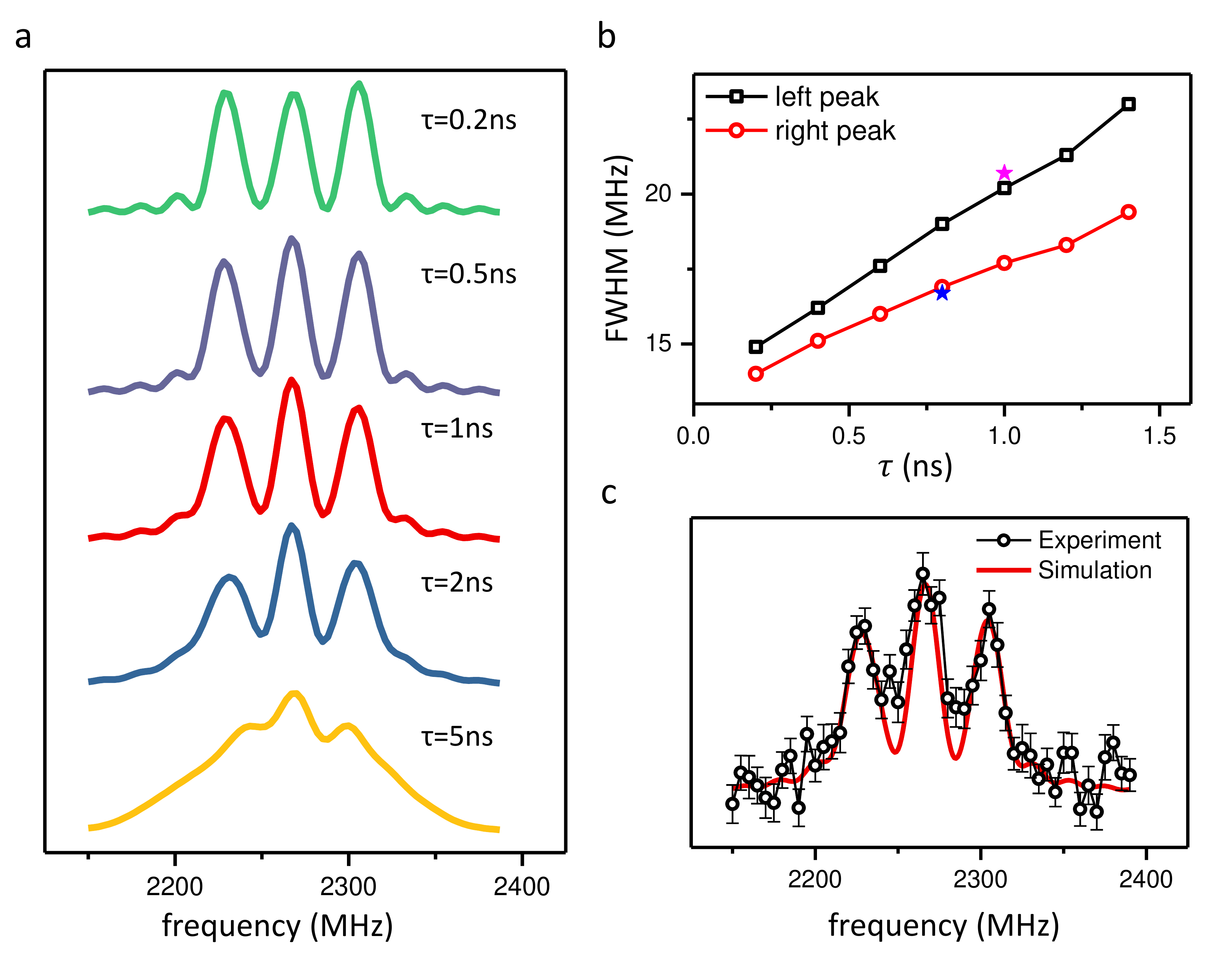} \protect\caption{\textbf{Spectral simulations.}
(\textbf{a}) Calculated spectra with different rotational correlation time $\tau$ as indicated.
(\textbf{b}) Plots of simulated linewidth versus $\tau$. The two stars mark the linewidth of left and right peaks of the NV-detected spectrum shown in Fig. 2a in the main text,  indicating that in the experimental spectrum, the motion of the spin label is best described by $\tau \sim1$ ns.
(\textbf{c}) Overlay of the NV-detected spectrum (black circles) with that simulated with $\tau=1$ ns (red line). The signal amplitude was scaled by a fix factor to best match the side-peaks. As shown the two side-peaks match well in both linewidth and amplitude between the experimental and simulated spectrum. The central peak of the experimental spectrum is clearly broader than the simulated one, reflecting the existence of background spin signals. The amplitude of the central peak also seems comparable between the experimental and simulated spectrum. This is possible because the existence of background spins does not always increase the signal contrast, as discussed in Supplementary Note 3E. Error bars indicate $\pm$1 s.e.m and the data shown were obtained from measurements repeated 0.8 million times.
}
\label{simulation}
\end{figure}

Supplementary Fig.~\ref{simulation}a shows spectra calculated with different rotational correlation time $\tau$. For fast rotations with $\tau=0.2$ ns or smaller, the calculated spectrum shows three narrow peaks with linewidth of $\sim$10 MHz, which is dominated by broadening due to the applied RF pulse. As rotations slow down to $\tau=1$ ns, the side peaks broaden to $\sim$20 MHz as observed in the NV-detected spectrum (Fig.~2a in the main text). Reducing the motion to $\tau=5$ ns leads to further peak broadening, and the spectrum starts to resemble that observed in the solid phase \cite{Du2015Science}.

As an example, we further estimated the value of $\tau$ for NV-detected spectrum shown in main text Figure 2a by comparing the experimental spectrum and the simulated ones (Supplementary Fig.~\ref{simulation}b $\&$ \ref{simulation}c). In these analyses, we matched only the linewidths of the side-peaks, which are not affected by the signal from diamond surface defects (Supplementary Note 3E). On a plot of simulated linewidth versus rotational correlation time $\tau$ (Supplementary Fig. \ref{simulation}b), the side-peak linewidths of the experimental spectrum best matched with those obtained from the simulation of $\tau=1$ ns. This is supported by direct overlay of the corresponding simulated and experimental spectrum (Supplementary Fig. \ref{simulation}c).

\clearpage

\begin{center}
\textbf{Supplementary References}
\end{center}
\renewcommand\refname{Supplementary References}